# Monolithic Integration of Sub-50 nm III–V Nano-Heterostructures on Si (001) for Telecom Photonics

Alisha Nanwani[1,2,*], Paweł Wyborski[1,3], Michael S. Seifner[2,4], Shima Kadkhodazadeh[2,4], Grzegorz Sęk[3], Kresten Yvind[1,2], Paweł Holewa[1,2,3,**], Elizaveta Semenova[1,2]

[1]Department of Electrical and Photonics Engineering, Technical University of Denmark, Denmark

[2]Nanophoton- Centre for Nanophotonics, Technical University of Denmark, Denmark

[3]Department of Experimental Physics, Faculty of Fundamental Problems of Technology, Wrocław University of Science and Technology, 50-370 Wrocław, Poland

[4]DTU Nanolab, Technical University of Denmark, Denmark

*atana@dtu.dk, **pawhol@dtu.dk

The demand for advanced photonics technology is increasing rapidly, fueled by the necessity for high-performance and cost-effective optical information processing systems extending into the quantum domain. Silicon, benefiting from its mature fabrication processes, stands as an ideal platform. However, its inherent indirect bandgap leads to inefficient light emission. The integration of III-V materials has been proven essential to overcome this drawback. These materials are recognized for their efficient light emission and superior bandgap engineering capabilities, making them indispensable in photonics and beyond. Here, we present the monolithic integration of small-volume III-V nano-heterostructures with silicon via selective area epitaxy in the pyramidal openings etched in (100)-oriented silicon substrate. The precise positioning of the nano-heterostructures is achieved using electron beam lithography. Our atomic resolution imaging and chemical analysis confirms the epitaxial nature of InP growth, revealing well-defined heterointerfaces. Each structure incorporates an InAsP quantum dot-like active medium, and the correlation of the growth parameters with the nanoscale structure was analyzed using advanced electron microscopy. The eight-band k·p calculations demonstrate energy level quantization in three spatial dimensions. Optical characterization shows that heterostructure emission can be engineered to cover the entire telecom wavelength range. Consequently, these InAsP/InP nano-heterostructures could serve as a gain medium for silicon-based hybrid nano-lasers and nano-LEDs and quantum light sources in the telecom wavelength range.

# Introduction

Integrating III-V active material with silicon platform has been a longstanding goal, leveraging the advantages of a well-established and cost-effective silicon photonic platform with addition of active photonic devices such as light sources and detectors[1–4]. In the past two decades, III-V/Si heterogeneous integration through bonding techniques has gained scientific and commercial recognition as a promising path towards realization of on-chip light sources [5–9]. With rapid advancements in the field of silicon-based integrated photonics for applications in artificial intelligence, hyper-scale data centers, high-performance computing, light detection and ranging (LIDAR)[10], monolithically integrated light sources are in high demand. The realization of a reliable method of monolithic integration of III-V compound semiconductor heterostructures onto host silicon-photonic circuits not only allows for higher integration density[11] but will also significantly simplify the technological process and decrease the overall emitter fabrication cost.

Monolithic integration of III-V compound semiconductors onto silicon via direct epitaxial growth has been extensively investigated for a few decades, with considerable progress occurring in recent years[12–20]. Various methods are actively explored in this respect, including the use of thick buffer layers[21–23], selective area growth (SAG) technique[24–33], the growth of nanostructures using droplets of metals as catalysts[34–36], a combination of SAG and metal droplet mediated approach[37], and aspect ratio trapping[38,39]. However, emitters fabricated using thick buffer layers can suffer from coupling issues with passive waveguides and recently some research groups have tried resolving this issue by etching the entire buried oxide (BOX) layer and regrowing thick buffers to compensate for the BOX.[40] Therefore, the latter three approaches are more promising due to their flexible integration into Si-based circuits, avoiding the use of expensive III-V substrates and wafer bonding process, as well as difficulties with the alignment of III-V elements with the Si-circuit.

Benefiting from high-quality III-V material growth obtained using these methods, various silicon-based laser structures have been demonstrated with outstanding performance[25,27,41,42]. SAG has emerged as a viable alternative, enabling precise positioning[27,43–45] of nanostructures with reduced defect densities due to a smaller footprint resulting in smaller nucleation area compared to the planar growth[44], and close to thermodynamic equilibrium growth conditions[34,36,46,47]. The accurate alignment of active III-V material with other photonic or electronic circuit layers is an enabling step for optoelectronic applications such as photonic cavities, nanolasers, or nano-LEDs.

Here, we demonstrate the realization of $InP/InAs_xP_{1-x}/InP$ nano-heterostructures of a few tens of nanometer size, monolithically grown on silicon and emitting in the telecom wavelength range. We chose InAsP gain material formed via As/P atoms exchange mechanism[48,49] over the growth of InGaAsP, conventionally used for emitters in telecom wavelength range, to mitigate the composition inhomogeneity caused by different migration lengths across different facets. We examine the controllability of the emission wavelength of the III-V nano-heterostructures by varying the arsenic content in the $InAs_xP_{1-x}$ region, exploring the potential applications of III-V nanostructures in silicon-based photonic devices. The nanostructures are investigated by scanning and transmission electron microscopy (SEM and TEM), revealing the geometry and composition of the InAsP quantum dot (QD)-like structures. The optical properties of nano-heterostructures are investigated by micro-photoluminescence (μPL) and time-resolved photoluminescence (TRPL) measurements. The results suggest that the nanostructures are supposed to be QD-like with the three-dimensional (3D) quantum confinement, further proven by the eight-band k·p calculations of the energy level structure.

# Results and discussion

## Epitaxial growth of III-V nano-heterostructures on silicon

Undoped silicon (001) wafers were used as a growth substrate, with e-beam patterned nano-sized openings varying in size from 20 to 50 nm. The pattern is transferred from a $SiN_x$ hard mask into the Si wafer by KOH etching, forming inverted pyramidal holes confined by the Si {111} surfaces[37]. The $InP/InAs_xP_{1-x}/InP$ nano-heterostructures are grown by the SAG method using a low-pressure (60 Torr) MOVPE reactor. Ultra-high purity hydrogen ($H_2$) is used as a carrier gas and trimethyl-indium (TMIn), phosphine ($PH_3$), arsine ($AsH_3$) and tertiary-butyl-phosphine (TBP) are the growth precursors (see Methods for description of the sample preparation and growth process). Figure 1a schematically illustrates the growth process, while Figs. 1b-d show the side and top view SEM images of the resulting nano-heterostructure with $InAs_xP_{1-x}$ QD.

First, the growth surface is prepared by heating up to 750°C under arsine/hydrogen ambient to remove the native oxide resulting in an expected (1x1) reconstructed surface. These (111) planes after surface preparation are composed of single atomic steps that help avoiding antiphase domains[33]. After surface preparation site-selective growth of an InP seed is optimized for different growth durations (40-110 s) at 600 °C with a molar flow ratio for the group V and III materials (V/III ratio) of 1690, with TMIn flux of $5.28 \times 10^{-6}$ mol/min and $PH_3$ flux of $8.92 \times 10^{-3}$ mol/min. InP growth time of 40 s was sufficient to fill the openings so that the active material is in-plane with the Si-circuit and also enough to trap the planar defects parallel to (111) planes at the $SiN_x$ undercut region. TMIn and $PH_3$ flows were paused for the QD-like structure formation, while $AsH_3$ was turned on to promote the As/P exchange. As a result, the nano-heterostructures exhibits InP {100}/InAsP {100} interfaces, which is confirmed by the high-resolution STEM images in Figure 2 and the schematics for the growth process are shown in Figure 1(a).

Table 1. summarizes the growth parameters used for three exemplary samples, S-1, S-2, and S-3. For sample S-1, the formation of the $InAs_xP_{1-x}$ layer was initiated by a 5-minute-long arsine exposure (As/P exchange) at 510 °C on an as-grown InP seed. It was subsequently capped with InP at V/III = 607 at the same temperature. The PL emission wavelength for this structure at low temperature (5 K) is 1240 nm (see the optical data

further in the text). Extending the arsine exposure to 7 min led to increased surface irregularities on the InP cap, indicative of defective growth, which was also observed in Ref. [50] with increased duration of As/P exchange. Consequently, instead of prolonging the arsine exposure, we increased the arsenic flux from $2.23 \times 10^{-3}$ mol/min to $7.81 \times 10^{-3}$ mol/min to redshift the emission wavelength to the C-band.

| Sample ID | InP seed growth time [s] | Annealing in $PH_3$ flux [s] | As/P exchange duration [s] | $AsH_3$ flux [mmol/min] | InP cap growth time [s] | Emission wavelength [nm] |
|---|---|---|---|---|---|---|
| **S-1** | 40 | 0 | 300 | 2.23 | 20 | 1240 |
| **S-2** | 110 | 800 | 180 | 7.81 | 20 | 1420 |
| **S-3** | 110 | 800 | 300 | 7.81 | 20 | 1620 |

*Table 1. Comparison of the growth parameters and PL emission wavelength for investigated samples.*

To further improve the crystal quality of the InP seed, we increased the growth duration to 110 s to overgrow the mask level, followed by annealing in $PH_3$ ambient for approximately 15 mins. Annealing promotes the desorption of the as-grown InP seed reducing its thickness to the same seed thickness as S-1 and regulates the defect count[37]. As/P exchange was then conducted for 180 s (S-2) and 300 s (S-3) at high arsenic fluxes, as shown in Table 1. We expect variations in thickness and composition of the $InAs_xP_{1-x}$ layer on the different crystallographic planes, due to crystal direction-dependent kinetics of the As/P exchange reaction.

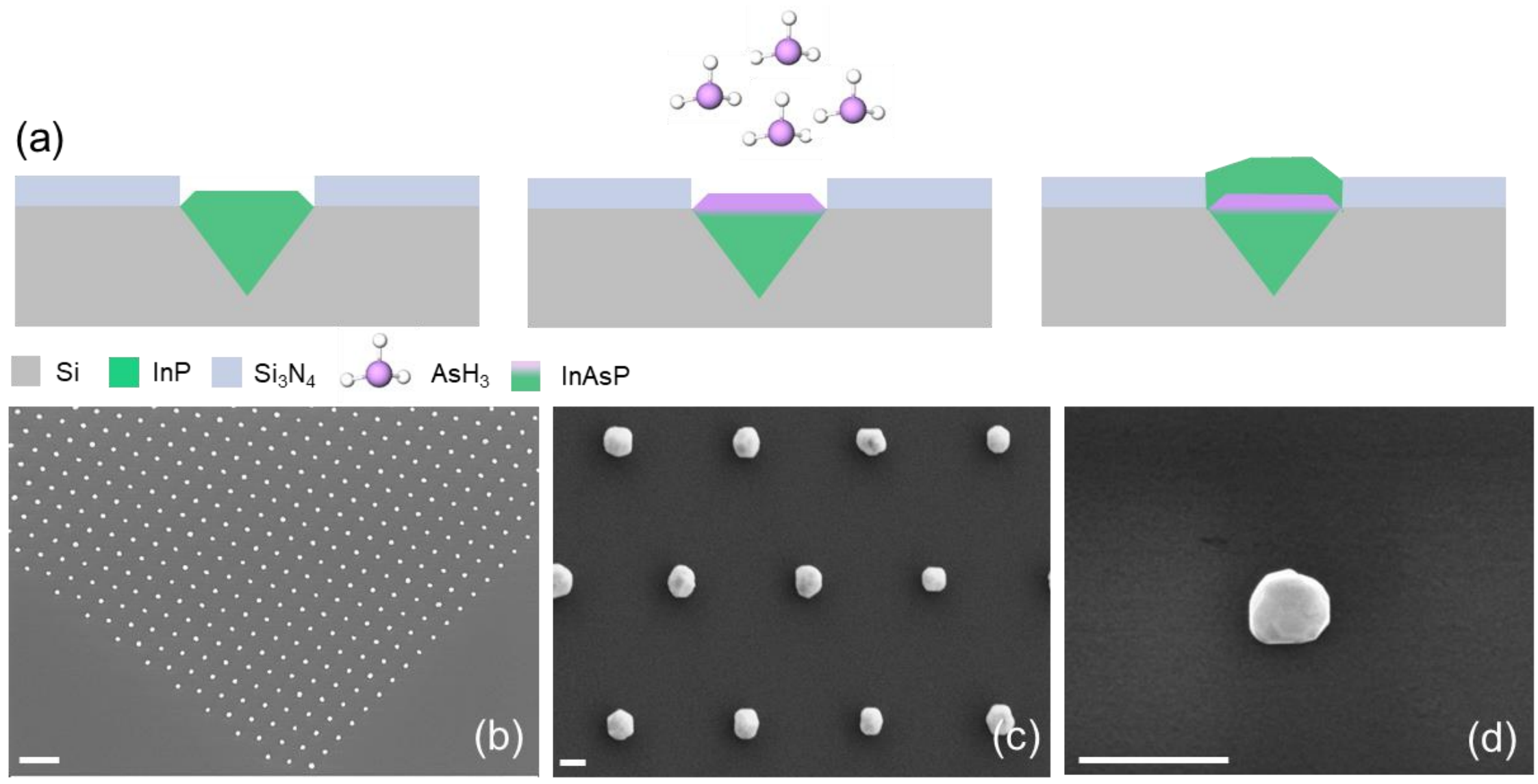

*Figure 1. Epitaxy of nano-heterostructures with InAsP quantum dot-like structure. (a) Schematic illustration of the growth procedure on silicon: the opening in Si is filled with the InP seed (left), the As/P exchange is performed in As ambient (center), the InAsP nano-heterostructure is capped with InP. Bottom row: SEM images of (b) 30×30 µm array of InAsP/InP nano-heterostructures grown, scale bar: 2 µm, (c) InP cap of the nano-heterostructures grown in 20 nm-size openings, scale bar: 200 nm, (d) a single 50 nm opening filled with a nano-heterostructure, scale bar: 500 nm.*

A long As/P exchange duration of 5 minutes results in an InAsP layer above pseudomorphic critical thickness that can be identified by a defective InP cap layer in S-3 (Supplementary Figure 4 and 5). The strain in the InAsP layer is most probably released via the formation of dislocations propagating in the InP cap layer, as also shown in Ref.

38, which results in the promotion of non-radiative carrier recombination channels. Also, longer exposure resulted in migration of indium atoms leaving voids in the seed layer shown in Supplementary Figure 5, locating the active medium above the silicon plane. The SEM images for as-grown nano-heterostructures are shown in Figure 1 (b-d); the InP cap shows non-uniform morphology; this is quite evident at a low growth temperature of 510°C as the migration length of indium is shorter, resulting in non-uniform growth across different facets[39].

## Nano-scale investigation of QD-like nano-heterostructures

Figure 2a presents a high-angle annular dark-field scanning transmission electron microscopy (HAADF STEM) image of an InP/InAsP/InP nano-heterostructure with a nominal 50 nm opening size for sample S-2. Supplementary Fig. 1 includes images and chemical analysis data for samples S-1 and S-3. High-magnification STEM images focusing on the As-containing heterointerface and the Si/InP interface are shown in Figs. 2b and 2c, respectively. Figure 2c highlights a sharp interface between Si and InP along the {111} boundaries of the etched pyramids (from S-3). Twins and stacking faults are known to be dominant defects for these interfaces. While twins are not detrimental to optical properties, the stacking faults are.  The anti-phase boundary (APB) formation on the Si (111) and (11-1) facets due to the merging of the nucleating crystals with the opposite phase was shown recently [51]. However, in our current study, we did not observe such APBs. The InAsP/InP QD-like structure can be clearly observed in Figure 2b and is confirmed by the EDS chemical composition maps shown in Figure 2e-g. The InAsP region has a thickness of 12 nm, and it is sandwiched between the InP seed and cap layers.

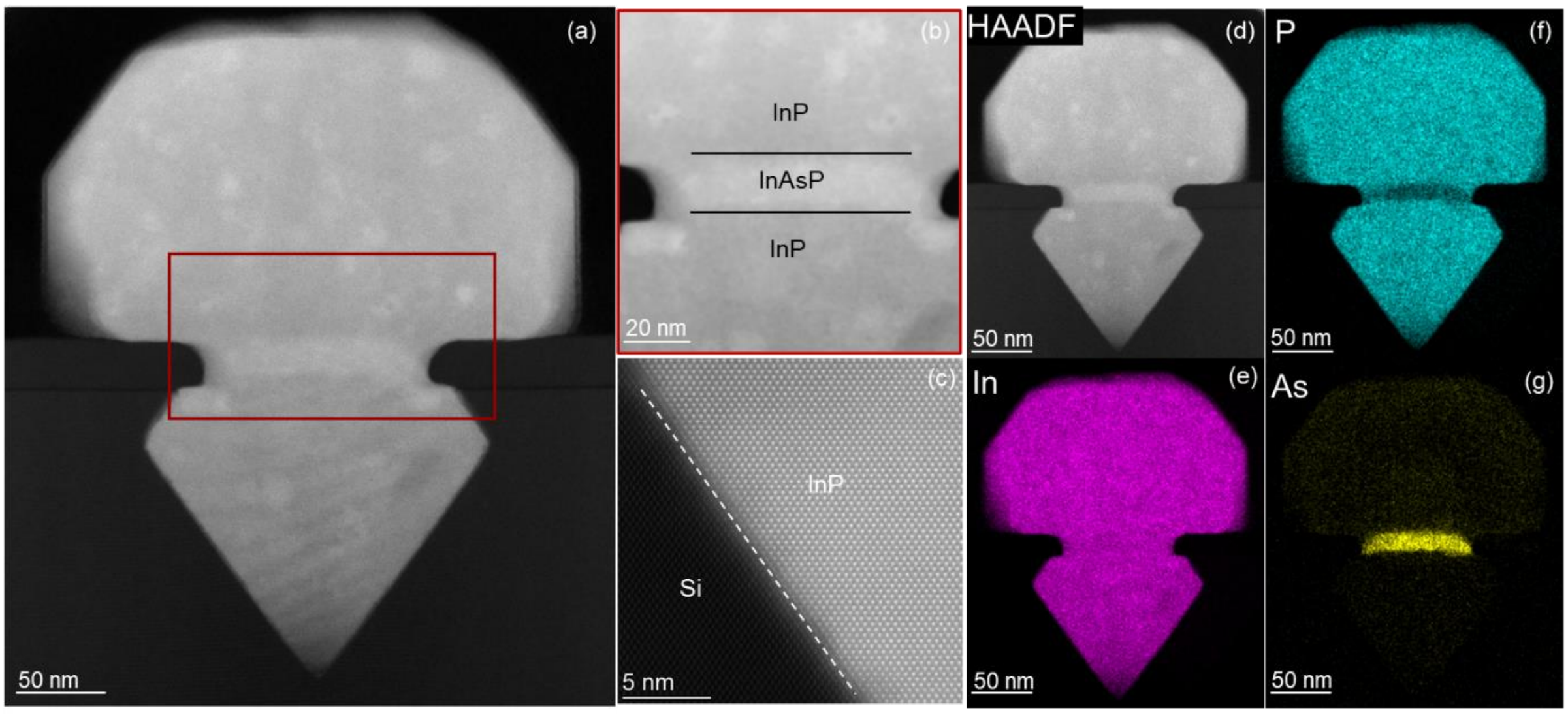

*Figure 2. STEM measurements of an InAsP QD-like nanostructure embedded in InP epitaxially grown on silicon of sample S-2, reveal, (a) an overview image and high-resolution images of (b) the InAsP and (c) Si/InP interface region from same batch of samples. (d)-(g) EDS analysis of InAsP/InP nano-heterostructure: (d) HAADF-STEM image, and (e-g) 2D elemental composition maps by EDS of (e) P (cyan), (f) In (magenta), and (g) As (yellow), respectively.*

Figure 2d-g presents the HAADF-STEM image and the corresponding elemental distribution of P (Fig. 2e, cyan), In (Fig. 2f, magenta), and As (Fig. 2g, yellow) along the heterointerface. It can be clearly seen that the distribution of In is uniform across the entire III-V heterostructure, whereas As appears only in the QD-like region, substituting the P sites to form $InAs_xP_{1-x}$ with an estimated average composition of x = 0.42. For sample S-1, the arsenic content was found to be x = 0.61. With the results of both STEM and EDS elemental mapping, it is therefore confirmed that the InP/InAsP interfaces are well-defined with good crystal quality.

## Optical properties of the InP/InAsP/InP/Si nano-heterostructures

The optical characterization of InP/InAsP/InP/Si nano-heterostructures was carried out in a μPL setup in the range of 5–250 K. The laser spot in the μPL experiment was approximately 2 μm, thus, it is expected to excite ~15-20 nanostructures (refer to Fig. 1c on geometry of the sample), averaging the optical response over a small ensemble. The results of the optical investigation are presented in Fig. 3.

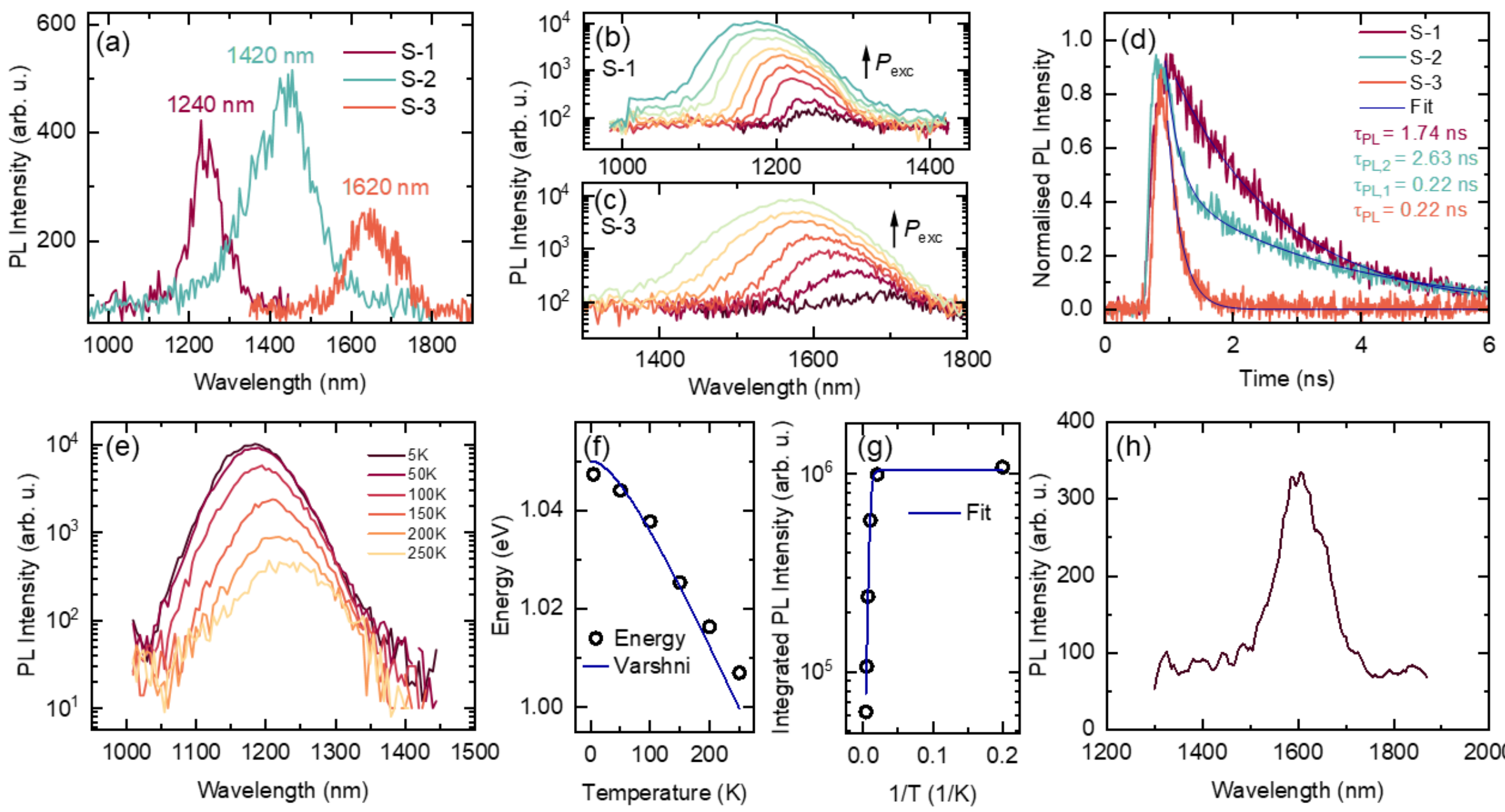

*Figure 3. μPL investigation of the QD-like structures at 5K, (a) μPL spectra taken from samples S-1, S-2, and S-3, (b), (c) excitation power dependent μPL spectra for samples (b) S-1, and (c) S-3, (d) time-resolved μPL traces for samples taken at their μPL peak wavelength, (e) temperature-dependent spectra for sample S-1, analyzed in (f) and (g): (f) emission energy overlapped with Varshni bandgap renormalization trend, (g) integrated μPL signal intensity with the Arrhenius fit line, (h) μPL spectrum of a single InP/InAsP/InP/Si structure recorded for sample S-3.*

All investigated samples emit in the telecom wavelength range: the optical emission is centered at 1240 nm for S-1, at 1420 nm for S-2, and at 1620 nm for S-3. The spectra taken at the 1 mW excitation power are plotted in Fig. 3a. The redshift of the emission wavelength correlates with the amount of arsine supplied during the As/P exchange, as shown in Table 1. The broadening of the PL peaks can be explained by the relatively high laser power needed for optical excitation (at least 20 μW), which increases the higher states occupation probability and hence their contribution to PL signal, thus increasing effectively the full width at half maximum (FWHM) of the PL peak. This behavior is caused by the low spacing of the quantized levels for electrons and holes, as described in the following section. Additionally, we expect the broadening due to the natural spectral diffusion originating in the fluctuating electric fields resulting from the accumulation and redistribution of charges either trapped in the vicinity of the InAsP nanostructure or at the InAsP/Si interface states. Consequently, the PL emission exhibits stochastic fluctuations in peak position and linewidth at timescales orders of magnitude lower than the integration time required to record a PL spectrum[52]. These fluctuations are thought to primarily influence the linewidth of the emission of single quantum emitters due to the inherent averaging of the emission energy that occurs during this process.

The excitation-power dependent µPL spectra for the nominally 50 nm-large openings from samples S-1 and S-3 are shown in Figs. 3b and 3c. As the laser power changes over three orders of magnitude from 20 µW to 20 mW, the central energy of the µPL peak shifts by 64 meV for S-1 and 52 meV for S-3 towards higher energies. The power-dependent blueshift of the emission energy accompanied by significant PL broadening increase can be explained by the state-filling effect in the QD-like structures: the finite density of states and the state energy quantization result in the increased probability of carrier population of the excited states. Consequently, the carrier recombination involving the excited states contributes to the increase of the PL intensity at the high-energy side of the emission peak (and hence also increases its width), effectively shifting the PL maximum to shorter wavelength. It is an expected observation in the excitation-power dependent PL spectra for nanostructures in which the confined states separation is smaller than the inhomogeneous broadening.

Subsequently, the time evolution of the µPL signal of InAsP QDs was examined via the TRPL measurements. Figure 3d shows the TRPL traces for all samples, taken at the energy corresponding to the maximum of their PL emission peak, at the power below the PL signal intensity saturation power. Therefore, we do not expect significant contribution of the excited states decay to the recorded signal. The TRPL trace for sample S-1 exhibits a monoexponential decay, which can be well described with the monoexponential function $I(t) = Ae^{-t/\tau}$, where $A$ is the scaling factor, and $\tau$ is the observed PL decay time. The fitted value of $\tau$ = 1.74 ns is in good agreement with the PL decay times typically observed for the InAs/InP QDs, grown in Stranski-Krastanov[53–55] or droplet epitaxy[56,57] mode, hence confirming the high structural and optical quality of S-1 structure, and being an indication of 3D quantum confinement. On the other hand, two decay time constants are needed to fit the TRPL trace for S-2, according to the bi-exponential decay formula $I(t) = Ae^{-t/\tau_1} + Be^{-t/\tau_2}$, with fitted values of $\tau_1$ = 0.22 ns and $\tau_2$ = 2.63 ns. The deviation from the monoexponential decay and the presence of an additional component can be explained by possible non-radiative carrier recombination channel. Among the possible non-radiative channels is the carrier relaxation using the mid-bandgap states, introduced into the InAsP bandgap by the structural defects in the crystal lattice or at the InAsP/Si interface. The TRPL decay time for S-3 is influenced even more by the non-radiative emission, demonstrating a relatively short monoexponential decay with the time constant $\tau$ = 0.22 ns. This can indicate the presence of a higher density of structural defects in the active material compared to the S-2, most probably originating from a larger volume of InAsP QD-like structure resulting from an excessively prolonged As/P exchange applied in sample S-3, shown also in Supplementary Figure S5.

The temperature dependence of the PL spectra has been analyzed to probe the thermal stability of emission and extract the carrier activation energies. Its example for sample S-1 is shown in Fig. 3e, while the analysis outcome is presented in Figs. 3f-g. The central emission energy follows the Varshni relation:[58]

$$E_g(T) = E_g(0) - \frac{\alpha T^2}{T + \beta}$$

where $E_g$ is the bandgap energy, and the parameters are taken for InAs, $\alpha = 2.76 \times 10^{-4}$ eV/K, and $\beta$ = 93 K. The corresponding curve is plotted atop the energies recorded for PL spectra in Fig. 3f. The empirical Varshni formula considers only the thermal change of the semiconductor band gap. The emission intensity decreases above 50 K, and this quench can be described by the Arrhenius formula that considers one activation process of energy $E_q$ and relative rate $B_q$[59]:

$$I(T) = \frac{I_0}{1 + B_q \exp\left(-E_q/k_B T\right)}$$

The obtained activation energy for the PL intensity decay is $E_q$ = (39.7 ± 5.6) meV. The calculated confinement energy for an exemplary $InAs_{0.45}P_{0.55}$ QD in the InP matrix is ~150 meV for electrons and ~250 meV for holes (see Fig. 4b in the next section). Thus, the PL intensity quench cannot be explained by a direct escape of carriers to the surrounding InP. The carriers might be excited to the mid-gap states introduced by the defects, where they recombine non-radiatively. Alternatively, as the band alignment for holes at the InAsP/Si interface can be of type II,[37,60–63] it might also be possible for the holes to escape from InAsP to the neighboring silicon matrix tunneling through a locally very thin InP barrier. This thin barrier can enable an interaction between the wavefunctions of the QD confined states and the InP/Si interface, which is expected to be defective.

The PL investigation was mainly performed for the nano-heterostructures with a surface density of 6 structures/$\mu m^2$. Therefore, the properties are averaged over a few emitters, and the observed broadening of the PL spectrum can partly result from the inhomogeneities of the emitter mini-ensemble or due to the spectral diffusion on the single-emitter level. In the first scenario, the As/P exchange efficiency depends on the orientation of the exposed InP facets, which are determined by the shape and size of the grown InP seeds. Any variation in this respect can introduce inhomogeneity into the size and composition of QDs within the excitation spot, thus broadening the PL peak.

Nevertheless, the PL investigations on a low-density part of the pattern allowed to record the spectra of single emitters. Their spectrum (shown in Fig. 3h) with an FWHM of 60 meV is similar to those recorded for the ensemble, favoring a second broadening scenario: in this case, the spectral diffusion on the single-emitter level has the decisive effect. The recorded FWHM value is comparable with the emission broadening for, e.g., CdSe/CdS/ZnS single colloidal quantum dots, where the spectral diffusion broadening mechanism was also observed[64]. We conclude that the FWHM of the ensemble PL peak at low excitation condition (when the contribution from the excited states PL is minimal) is determined

## Calculations of the electronic structure

We calculated the electronic structure of the $InAs_xP_{1-x}$ QD-like structures in the InP matrix within the eight-band k·p method to investigate the carrier confinement and the agreement between the expected electron-hole (e-h) recombination energy with the EDS data. We employed the commercially available nextnano software[65] to calculate the single-particle states for both electrons and holes. For this, the continuum elasticity theory is utilized to model the strain distribution, followed by the calculations of the carrier eigenstates within the eight-band k·p method, including the strain-driven piezoelectric field. The results of these calculations are shown in Fig. 4. Excitonic effects are not accounted for, however they were not necessary to confirm the quantum size effect and would not be deciding about the interpretation of the experimental data at the observed inhomogeneities.

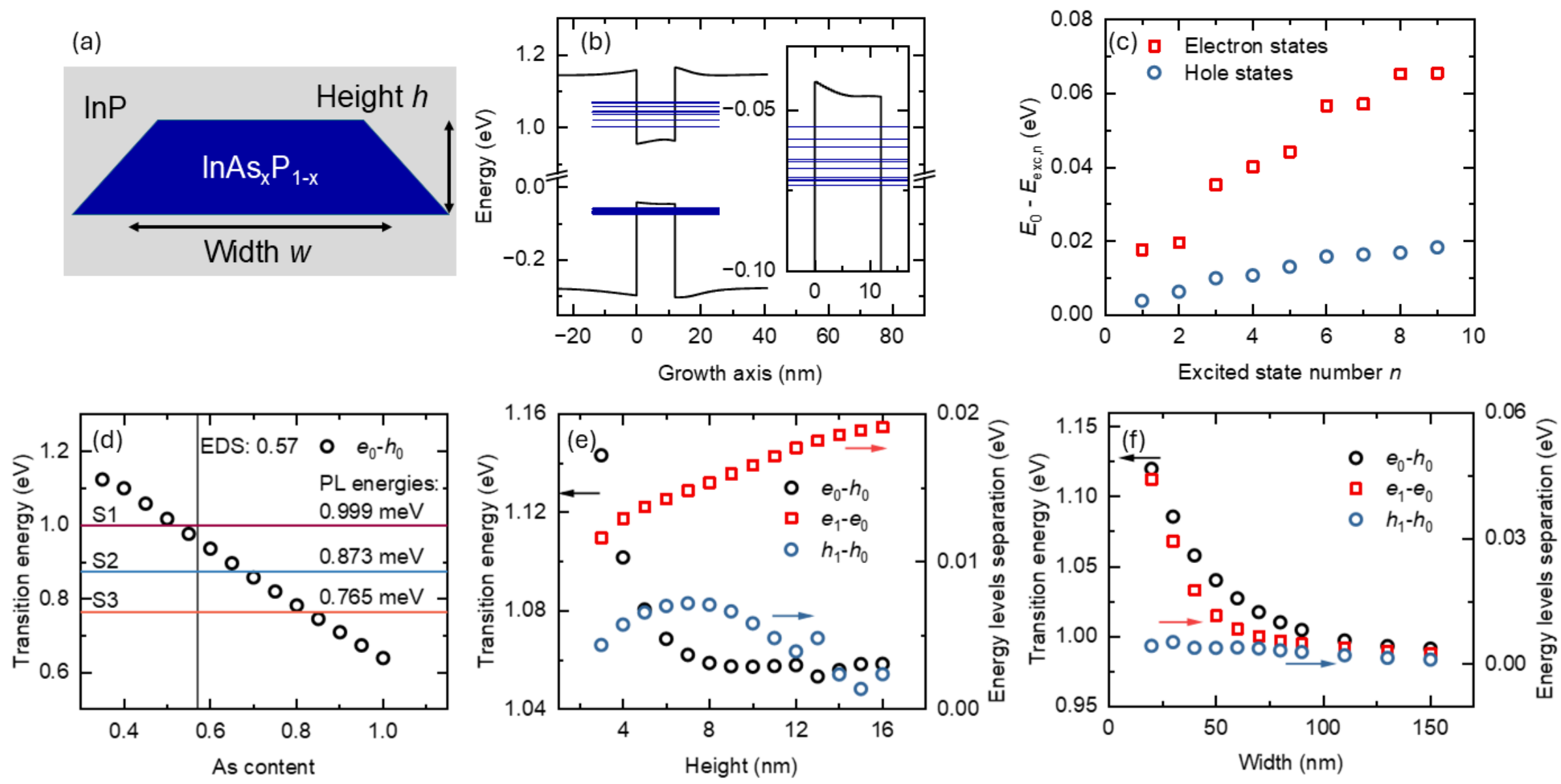

*Figure 4. Calculations of the InAsP/InP electronic structure in eight-band k·p. (a) Scheme of the investigated structure, (b) conduction and heavy hole valence bands with electron and hole states confined in the InAsP QD-like structure (inset: close up to the valence band), (c) energy levels separation between the ground state $E_0$ and consecutive excited states $E_{exc,n}$ - absolute value, (d), (e), (f) transition energy between the ground electron and heavy-hole states $e_0$-$h_0$ (black datapoints) as a function of (d) As content in the nanostructure, overlaid with the PL peak energies for all three samples, and the EDS composition of $x = 0.57$ for S-1 indicated by the vertical line, (e) nanostructure's height, and (f) its width. In panels (e) and (f), the*

*right axis is used to show the separation between the ground and first excited states for electrons ($e_1$-$e_0$, red squares) and holes ($h_1$-$h_0$, blue circles).*

Fig. 4a shows the scheme of the simulated structure, with the geometry based on the STEM image shown in Fig. 2. We model the nanostructure as an in-plane symmetric $InAs_xP_{1-x}$ truncated pyramid embedded in the InP matrix. We assume the wall inclination angle of 45 °, though it is impossible to infer the full 3D model of the $InAs_xP_{1-x}$ from the 2D projection of the sample obtained from STEM. However, this is not required, as our motivation is to describe the regime of confinement and the general trends. We take the dimensions and composition of the nano-heterostructure as determined in the STEM images shown in Fig. 2: height of 12 nm, width of 40 nm, and the As content of x = 0.45, though the width and height might deviate across the ensemble of emitters.

The conduction and heavy-hole valence bands taken along the growth axis at the center of the structure are shown in Fig. 4b, together with energies of the first 9 hole and 9 electron confined states. The pronounced energy difference between the ground states and the respective $InAs_{0.45}P_{0.55}$ band edges, together with the clear discrete state energy ladders, evidence the quantum confinement in the nanostructures. Supplementary Figure 2 shows the calculated probability distributions for ground and first excited states. Figure 4c shows the energy difference between the ground and consecutive excited states for holes and electrons. For the first excited electron and hole states, the energy separation from the respective ground states is 17.7 meV and 3.9 meV, respectively. The relatively dense energy ladders for electrons and holes support the interpretation that in the regime of high excitation power, many excited states have a relatively high occupation probability, broadening the PL spectrum.

We define the transition energy as the difference between the ground state levels for electrons and holes, $e_0$-$h_0$, and show its dependence on the As content in Fig. 4d, on the QD height in Fig. 4e, and on its base width in Fig. 4e. All parameters influence the transition energy, with As content having the largest influence (energy change from 1.13 eV to 0.63 eV when the As composition is increased from 35 % to 100 %). We found that at least 35 % of As in the nanostructure is required to confine both carriers in the InAsP potential. The expected transition energy for the InAsP nanostructures with As content of x = 0.57 (sample S-1) is 0.96 eV, which is in good agreement with the central PL energy of 0.99 eV, recorded for the same sample.

Based on the calculations of the energy levels for different heights, shown in Fig. 4e, the quantum size effect along the growth direction is most pronounced in the structures thinner than 10 nm, though the quantization energy (the energy difference between the first excited state and the ground state) in both bands depends on the size in the entire investigated range of 3-16 nm. We note that the InAsP nanostructure's height of 12 nm, as determined by the STEM, can be easily overestimated due to intermixing of As and P atoms.

Importantly, the three-dimensional confinement of carriers can be confirmed by the evolution of the calculated energies for varying InAsP width, plotted in Fig. 4f. The quantum size effect can be observed as a change in the transition energy occurring for nanostructures narrower than 110 nm. Based on the changes in the quantization energies for electrons and holes (right axis in Fig. 4f), we can conclude that the electron states ladder is much more affected by the change of the in-plane size of the nanostructure due to the smaller electron effective mass. The calculated changes in the transition and the quantization energies for the width range realized in investigated nanostructures confirm the existence of the in-plane quantum confinement, and in consequence, of the three-dimensional quantum confinement. This observation further supports terming the InAsP nanostructures as quantum dots. The small energy level separation for carriers, which is less than the exciton binding energy for QDs of similar sizes[66,67], demonstrates that the InAsP/InP QDs investigated here represent the weak carrier confinement regime.

# Conclusions

In conclusion, we demonstrated the monolithic integration of the InAsP/InP nano-heterostructures on silicon. The nano-heterostructures are grown in the pyramidal openings terminated by the {111} surfaces of the (100)-oriented silicon substrate. The positions of the nanostructures are predetermined by the e-beam lithography and selective anisotropic wet etching of silicon. The systematic investigation of the morphology, chemical composition, and optical properties of the nano-heterostructures allowed the evaluation of their geometry and revealed their QD-like behavior. Moreover, the eight-band k·p calculations based on morphological studies prove the energy levels' quantization in three spatial dimensions, even though the applied high excitation power significantly broadened the optical response of the nanostructures by exciting higher order states, together with the effect of the existing significant spectral diffusion. We demonstrated the optical activity of such dots in the telecom range, while their spectral characteristics indicate that there is still room for improvement in their performance by limiting the influence of the defect states in the InP surrounding.

Our results show great potential for fabricating light sources grown directly on silicon and emitting at various wavelengths covering the entire telecom range for diverse optical and silicon photonic-based applications, such as CMOS-compatible energy-efficient nano-lasers or nanoLED with a nanometer-scale active region. Future optimization should focus on engineering the photonic environment of the nano-heterostructures to increase the light outcoupling efficiency.

# Methods

## Substrate preparation

Samples were prepared by starting with an undoped Si (001) wafer, on which 23 nm-thick $SiN_x$ was deposited by LPCVD. The wafer was patterned using e-beam lithography in a JEOL JBX-9500FSZ e-beam writer using the 100 kV acceleration voltage and CSAR (AR-6200.09) resist. After developing in ZED N-50 for 1 min, the $SiN_x$ was dry etched in ICP-RIE using $CHF_3/CH_4$ chemistry. Oxygen plasma ashing was used to strip the resist, followed by KOH etching to form inverted pyramids in silicon at 70 °C. Piranha cleaning was performed before dipping the sample for 3 mins into a buffered HF solution and immediately loading it into the MOVPE chamber to prevent oxidation of the silicon surface.

## Epitaxial growth

The growth was carried out in an MOVPE reactor (Emcore D-125 Turbo-disc) at 60 Torr (1000 rpm) using 22.4 l/min of hydrogen as the carrier gas. TMIn, $PH_3$, and $AsH_3$ were used as precursors for In, P, and As, respectively. First, the surface preparation was performed at 750 °C under the high $AsH_3$ flux. Then, the temperature was ramped down to the growth temperature of 600 °C, and TMIn and $PH_3$ were simultaneously turned on to start the InP growth. Then, InAsP nanostructures were formed by shutting off TMIn and cooling down to the exchange temperature of 510 °C under TBP flow, followed by turning on $AsH_3$, and simultaneously turning off TBP flows. For the growth of the InP cap layer, the temperature is kept at 510 °C. After the growth, the reactor is cooled down with TBP flow to prevent the desorption of the cap layer.

## Optical characterization

The samples with the nano-heterostructures were held in a helium-flow cryostat, allowing for control of the sample temperature in the range of 4.2-300 K. For the standard μPL studies, the structures were optically excited through a microscope objective with NA = 0.4 and 20× magnification using 640 nm or 805 nm light generated with semiconductor laser diodes. The same objective was used to collect the μPL signal and to direct it for spectral analysis into a 0.32 m-focal-length monochromator equipped with a 600 grooves/mm diffraction grating and a liquid-nitrogen-cooled InGaAs multichannel array detector, providing spatial and spectral resolution of about 2 μm and 100 μeV, respectively.

The time-resolved μPL was measured in the same setup. Here, the nanostructures were excited by 50 ps-long pulses with a repetition rate of 80 MHz and a central wavelength of 805 nm. At the same time, the second monochromator output port was equipped with the fiber coupling system, transmitting the signal to an NbN-based SNSPD (Scontel) with above 50 % quantum efficiency in the range of 1.2-1.6 μm and 100 dark counts per second. A multichannel picosecond event timer (PicoHarp 300 by PicoQuant GmbH) analyzed the single photon counts as a time-to-amplitude converter. The overall time resolution of the setup is estimated to be 80 ps.

## Lamella preparation

The lamellae were prepared with a plasma focused ion beam scanning electron microscope (plasma FIB-SEM, Thermo Fisher's Helios 5 Hydra Ux DualBeam). SEM images were acquired with accelerating voltages between 2 kV-5 kV and electron currents between 100 pA-200 pA. Secondary electrons were used as signal. The field-free mode was chosen for sample navigation, while for high-quality imaging the immersion mode was applied. Xe ions were used for sample navigation with the ion beam (accelerating voltage: 30 kV, ion current: 10 pA). The electron beam was used to deposit a thin layer of carbon followed by a slightly thicker layer of a carbon/platinum mixture (accelerating voltage: 2 kV, electron current: 1.6 nA). Subsequently, a thick layer of a carbon/platinum mixture was deposited by the ion beam (accelerating voltage: 12 kV, ion current: 1 nA).

The lamellae were prepared by milling trenches (regular cross-section, accelerating voltage: 30 kV, ion current: 60 nA), followed by cleaning the surfaces (cleaning cross-section, accelerating voltage: 30 kV, ion current: 40 nA) and a J-cut (accelerating voltage: 30 kV, ion current: 15 nA). After a final cleaning of the surfaces, the lamellae were attached to a needle (Pt/C, accelerating voltage: 30 kV, ion current: 100 pA), the remaining bridge was cut (accelerating voltage: 30 kV, ion current: 1 nA), and the structure was lifted out of the substrate. Finally, the lamellae were attached to a TEM grid suitable for electron microscopy investigations, followed by precise thinning (Xe, accelerating voltage: 30 kV, ion current: 200 pA) and polishing steps (Ar, accelerating voltage: 5 kV/2 kV, ion current: 61 pA/25 pA) to make the quantum dots accessible for STEM measurements.

### Electron microscopy

A probe and image-corrected state-of-the-art electron microscope (Thermo Fisher's Ultra Spectra STEM) was used for the study. The microscope is equipped with an Ultra-X EDS system for the efficient collection of EDS data. HAADF STEM images were acquired with an accelerating voltage of 300 kV after probe correction with convergence angles between 30 mrad and 18.5 mrad and a camera length of 109 mm. The HAADF STEM and EDS data were processed and analyzed using Velox from Thermo Fisher Scientific (Version 3.14.0) and DigitalMicrograph from Gatan (Version 3.60.4435.0).

## Acknowledgments

We acknowledge financial support from the Danish National Research Foundation through NanoPhoton - Center for Nanophotonics, grant number DNRF147, and the National Science Centre in Poland within grant No. 2019/33/B/ST5/02941. We are thankful to Michał Gawełczyk from Wroclaw University of Science and Technology for fruitful discussions and valuable suggestions concerning quantum dot electronic structure calculations.

## Competing interests

The authors declare no competing interests.

## Data Availability Statement

The authors confirm that the data supporting the findings of this study are available within the article and its supplementary material.

# Supplementary Information

# Monolithic Integration of Sub-50 nm III–V Nano-Heterostructures on Si (001) for Telecom Photonics

Alisha Nanwani[1,2,*], Paweł Wyborski[1,3], Michael S. Seifner[2,4], Shima Kadkhodazadeh[2,4], Grzegorz Sęk[3], Kresten Yvind[1,2], Paweł Holewa[1,2,3,**], Elizaveta Semenova[1,2]

[1]Department of Electrical and Photonics Engineering, Technical University of Denmark, Denmark

[2]Nanophoton- Centre for Nanophotonics, Technical University of Denmark, Denmark

[3]Department of Experimental Physics, Faculty of Fundamental Problems of Technology, Wroclaw University of Science and Technology, 50-370 Wroclaw, Poland

[4]DTU Nanolab, Technical University of Denmark, Denmark

*atana@dtu.dk, **pawhol@dtu.dk

## S1 EDS Analysis for samples S-1 and S-3

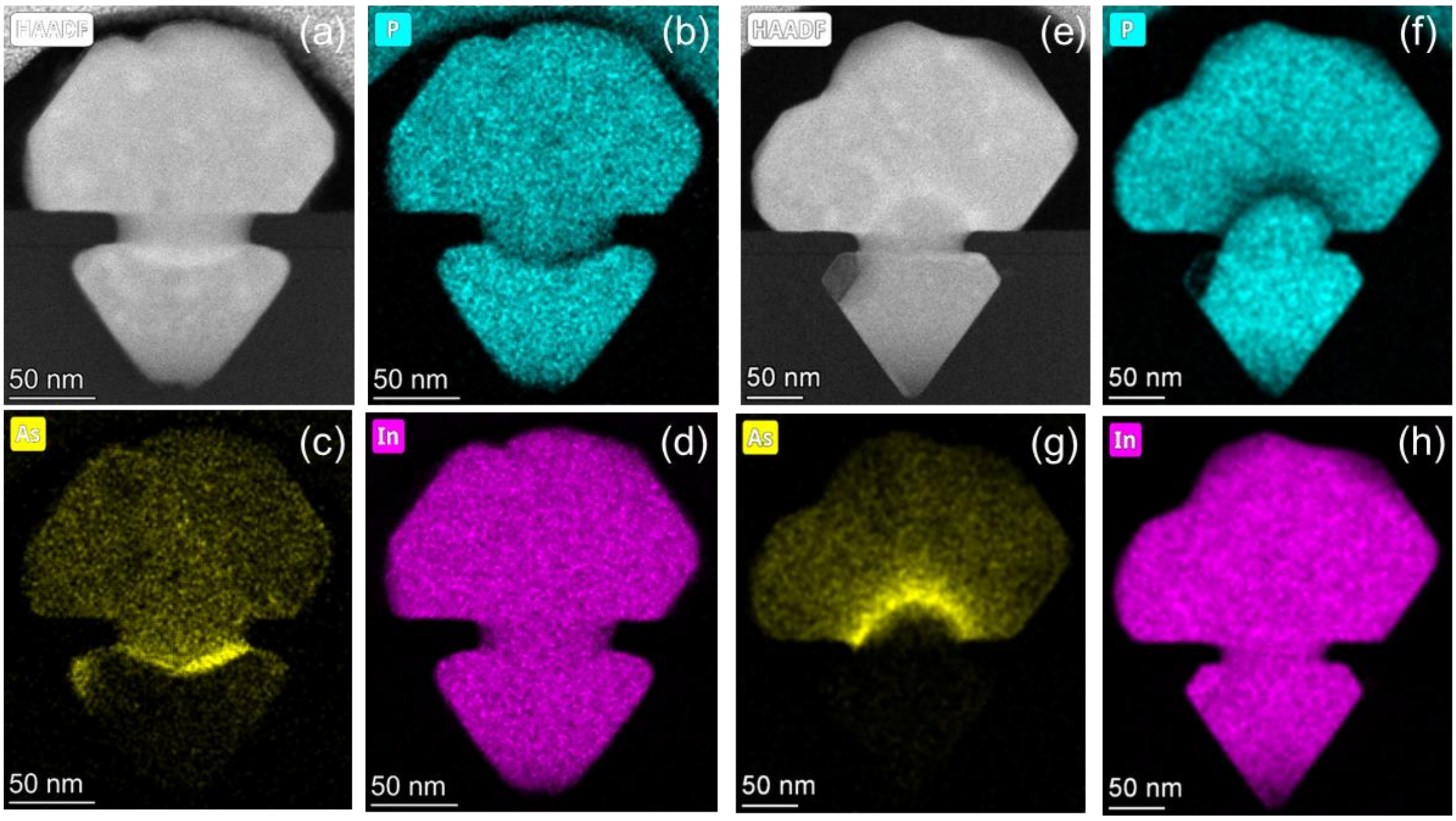

*Supplementary Figure S1. EDS analysis of InAsP/InP nano-heterostructures, S-1 (left) and S-3 (right): (a) HAADF-STEM image for S-1, and (b-d) 2D elemental mapping of S-1 (b) P (cyan), (c) As (yellow) and (d) In (magenta), respectively. (e) HAADF-STEM image for S-3, and (f-h) 2D elemental mapping of S-3 (f) P (cyan), (g) As (yellow) and (h) In (magenta), respectively.*

## S2 Calculated probability distributions

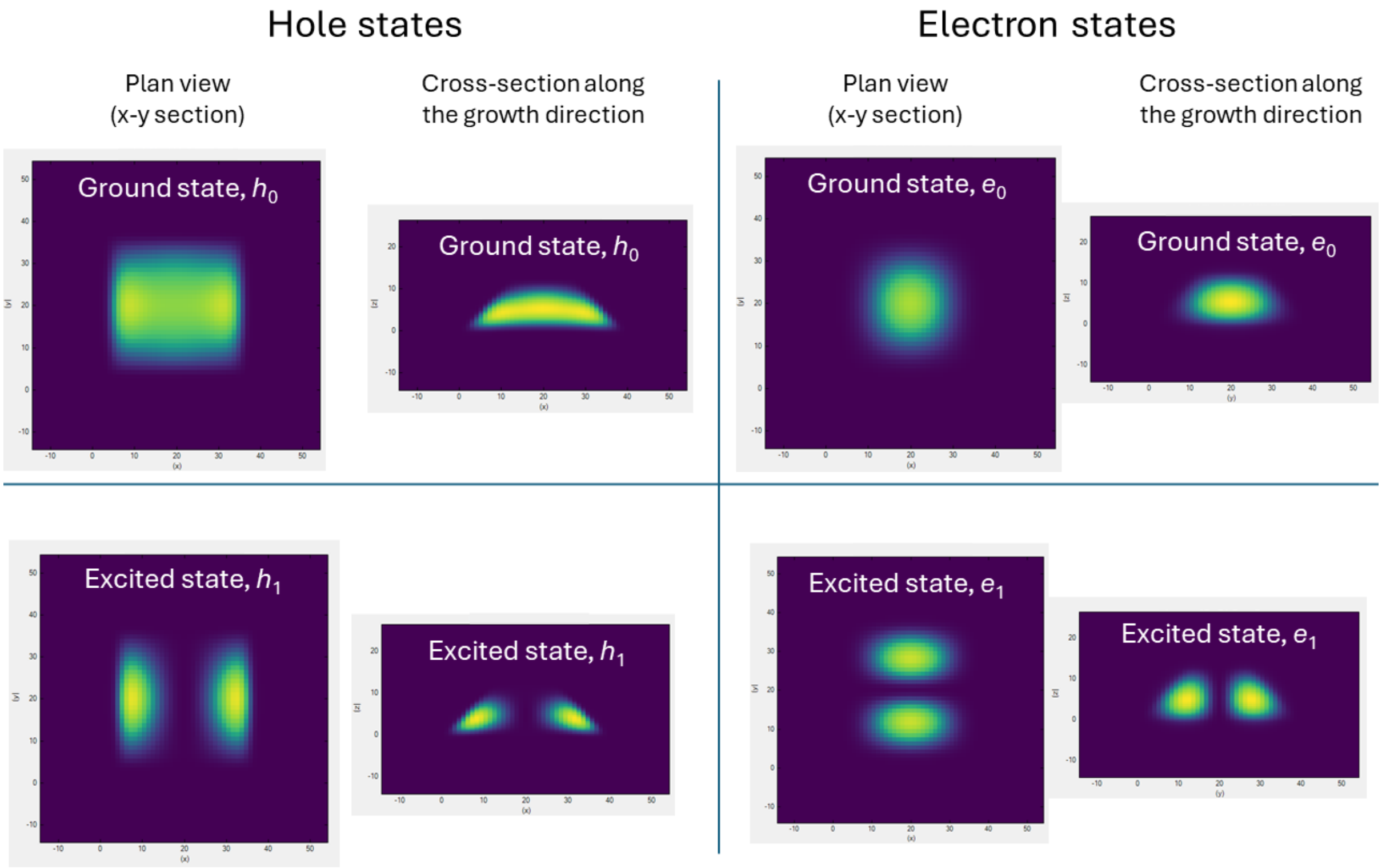

*Supplementary Figure S2. Calculated probability distributions for the hole (h, left) and electron (e, right) states: ground (top row, $e_0$, $h_0$) and the first excited state (bottom row, $e_1$, $h_1$), for the $InAs_{0.45}P_{0.55}$/InP nano-heterostructure.*

S3 EDS line scans for sample S-2

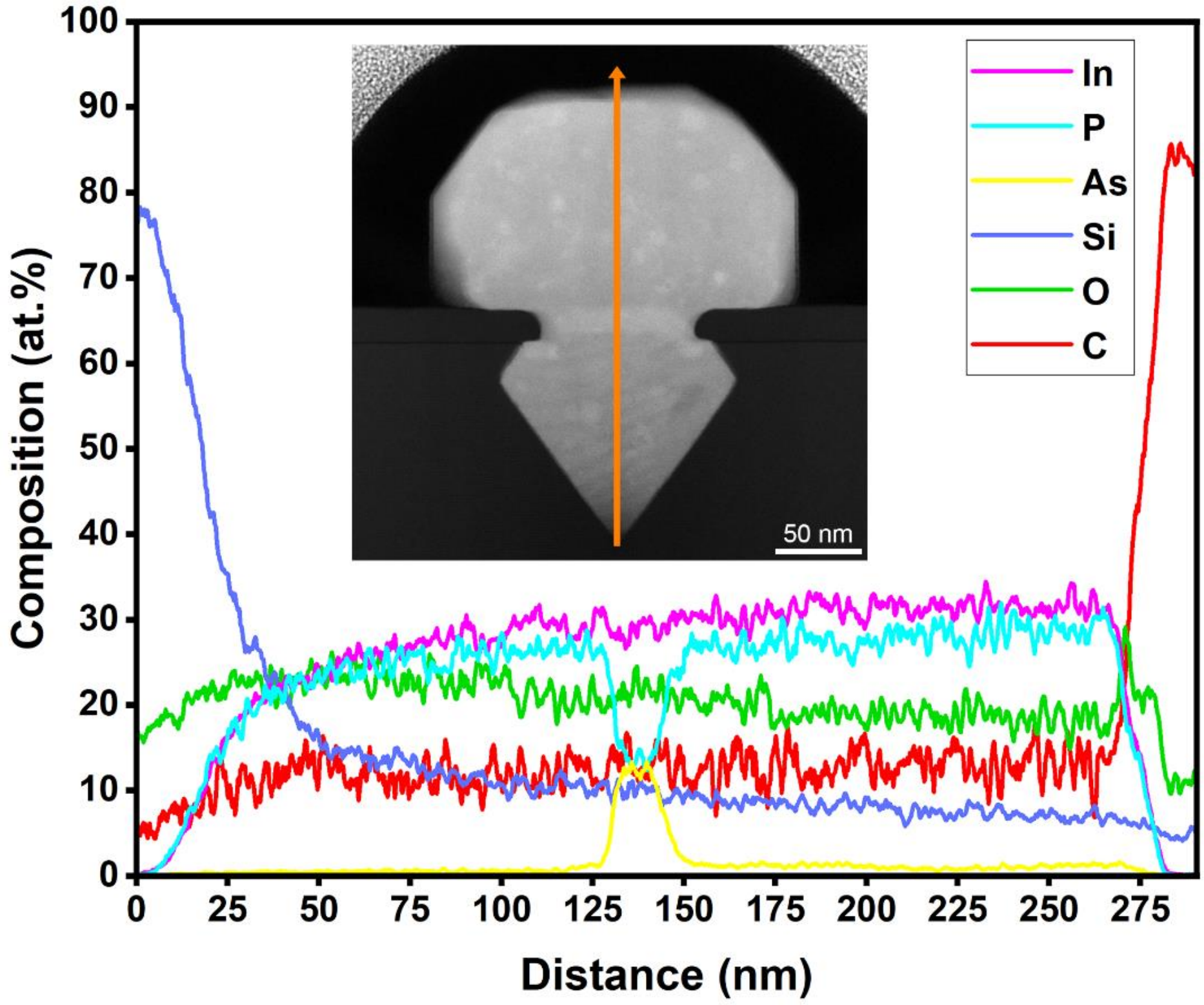

*Supplementary Figure S3. 1D line scans by EDS showing the atomic concentration of elements As, P, In, in the heterostructure along the growth direction. The inset is HAADF image for nano-heterostructure with line marked along which line scans are measured.*

## S4 SEM images for S-2 and S-3

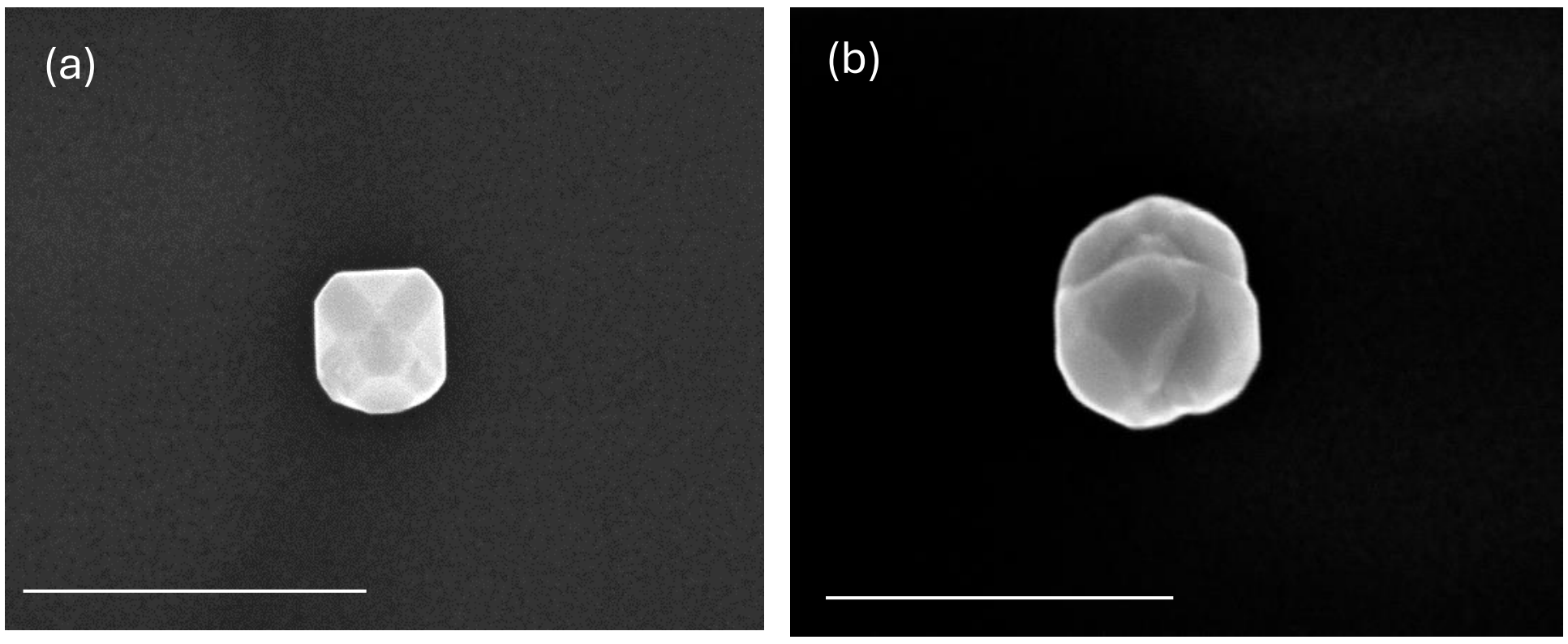

*Supplementary Figure S4. Top view SEM images for 50 nm opening size (a) S-2, (b) S-3, the defective cap for S-3 clearly shows defects propagating towards the surface. Scale bar: 500nm*

## S5 STEM images S-3

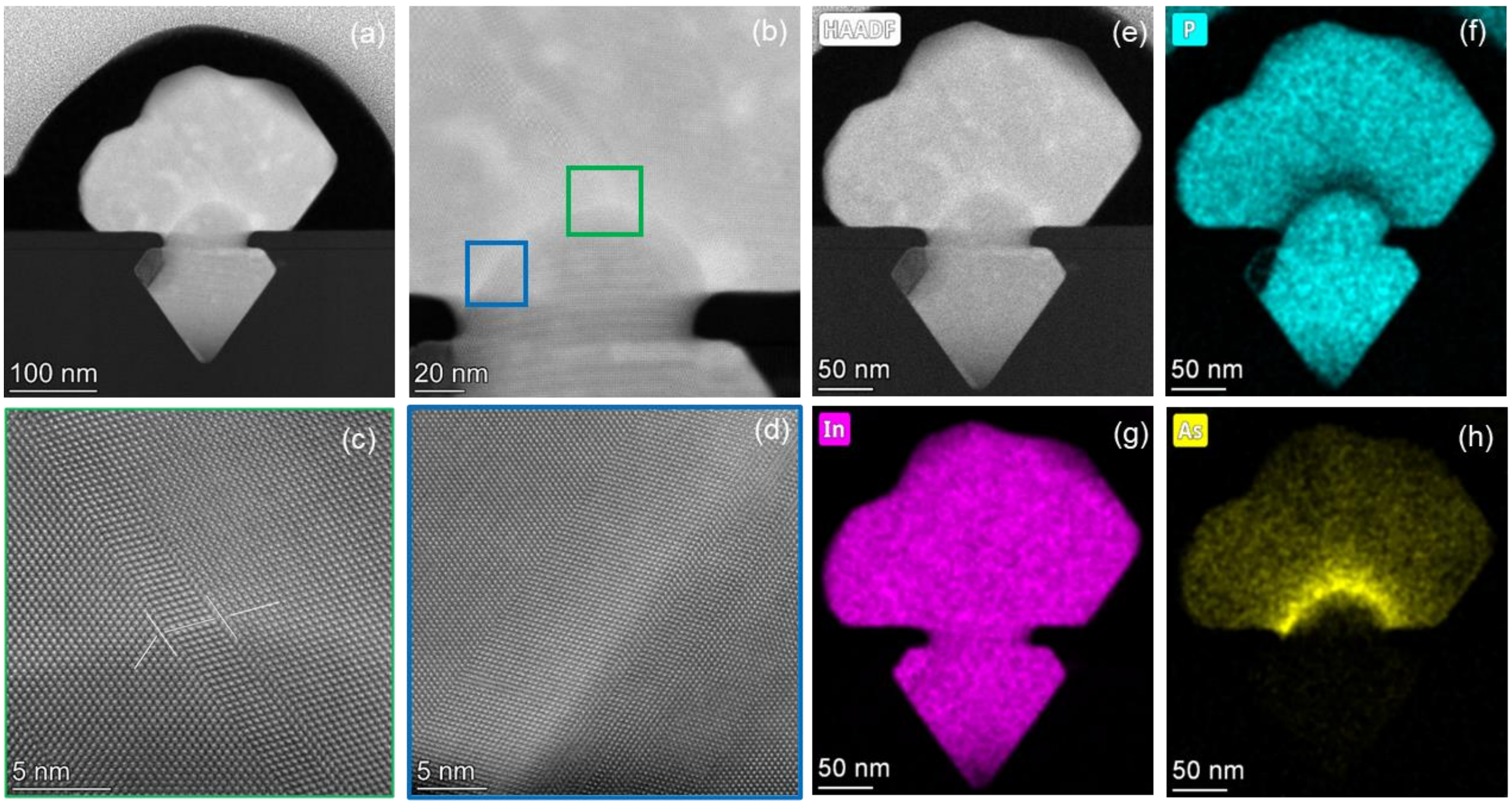

*Supplementary Figure S5. STEM measurements of an InAsP QD-like nanostructure embedded in InP epitaxially grown on silicon of sample S-3, reveal, (a) an overview image and high-resolution images of (b) the InAsP and (c) the planar defect region propagating to the surface (d) atomic resolution of QD region. (e)-(g) EDS analysis of InAsP/InP nano-heterostructure: (e) HAADF-STEM image, and (f-h) 2D elemental composition maps by EDS of (f) P (cyan), (g) In (magenta), and (h) As (yellow), respectively.*